# Optimal Voltage Control using Singular Value Decomposition of Fast Decoupled Load Flow Jacobian

Talha Iqbal, Ali Dehghan Banadaki, Ali Feliachi
Lane Department of Computer Science and Electrical Engineering
West Virginia University
Morgantown, WV 26505-6109, USA
Email - ti0001@mix.wvu.edu

*Abstract*—The problem of regulating voltages within the required limits is complicated by the fact that power system supplies power to a vast number of loads and is fed from many generating units. As loads vary, reactive power requirements of the transmission system vary. Moreover, voltage magnitude is relatively less sensitive to active power compared to reactive power due to high X/R ratio of transmission lines. Therefore, separating voltage control from active power is not only justified but also the common and practical way in power transmission systems. Considering these facts, the fast decoupled power flow jacobian can be used to control voltage magnitudes by reactive power compensation. In this paper, an optimal voltage control is presented to obtain new voltage set-points for PV buses by maximizing the effect of input change on output change using the Fast Decoupled Load Flow (FDLF) jacobian matrix. The proposed algorithm was tested on three IEEE systems: 9 bus, 14 bus and 30 bus systems.

## I. Introduction

Voltage and frequency control in power systems have always been considered as two fundamental regulation problems. Frequency regulation through active power control was considered and settled first but voltage control problem has no standard solution yet [1]. Voltage regulation through active power control is also theoretically possible but this method is never used in practice except under extreme operating conditions where there are high system security risks [2-4]. In modern power networks, voltage and frequency control with optimal operation of the network is a big challenge. Voltage regulation in power system is complicated by the fact that it supplies power to a large number of loads and has many generating units including renewable energy resources [5-9]. As loading conditions are changed, the reactive power requirements of the transmission system vary [10]. Moreover, in transmission systems, due to high X/R ratio of transmission lines, voltage magnitude is less sensitive to active power and relatively more sensitive to reactive power. Therefore, decoupling voltage control from active power is not only justified but also the common and practical way in power system operations [11]. Considering these facts, fast decoupled load flow assumptions can be used for voltage control.

In transmission systems with X >> R, voltages can be controlled by the injection or absorption of reactive power. In general, five methods of injecting reactive power are available: static shunt capacitors, static series capacitors, synchronous compensators, static VAR compensators and STATCOMs [2]. Currently, the system voltage profile is kept within normal operating limits by putting a reactive power source at the bus, changing transformer tap ratio or controlling the generator terminal voltage [12]. Voltage control based on sensitivity analysis has been a hot research topic for the last few decades. A voltage control technique based on defining voltage control areas using the jacobian matrix is presented in [13]. But this method has a high degree of trial and error [14-15]. [16] discusses a method of controlling voltage based on the structure of the network. Voltage control areas are determined based on electrical distance between the buses. This method of finding electrical distance has been applied in [17] and [18]. In [19], voltage control areas are identified based on direct relationship between generator's reactive power and load, by finding a sensitivity matrix that relates the reactive power output of a generator to a load.

In this paper, an optimal voltage control algorithm is proposed to obtain new voltage set-points for PV buses by maximizing the effect of change in input (i.e. $\Delta V_{PV}$) on output change (i.e. $\Delta V_{PQ}$) using the FDLF jacobian matrix, so that we can have maximum reactive reserve and minimum shift of the controls. The remainder of the paper is organized as follows: The proposed optimal voltage control approach is formulated in Section II. Section III describes all steps involved in the proposed voltage control algorithm through a control flow diagram. Section IV discusses implementation and simulation results of the proposed voltage control approach through case studies. Performance comparison with sensitivity based voltage control approach is presented in Section V. Conclusions and discussion on future work are discussed in Section VI.

## II. Problem Formulation

### A. Eigenvalues and Eigenvectors

Linear systems can be represented mathematically by linear matrix equations. Consider a linear system represented by a 2x2 matrix *A*. It takes *x* as input vector and transforms it linearly into *y* as output vector as shown in Fig. 1, where *y* = *Ax*. When *A* operates on *x*, it changes its magnitude as well as direction, however there are some special vectors which, when applied on the system, do not change their direction but only magnitude is affected i.e. they are stretched or squeezed only. These special vectors are called *eigenvectors* ($v_\lambda$) of the matrix $A$ and the amount of stretch or squeeze is called its *eigenvalue* ($\lambda$). In other words, when we apply $v_\lambda$ (an eigenvector of $A$) to $A$, we get another vector whose direction is the same as $v_\lambda$ but its magnitude is scaled by some factor which is called eigenvalue i.e. eigenvectors are those vectors which do not knock off their span when operated by the linear system $A$ [20-21].

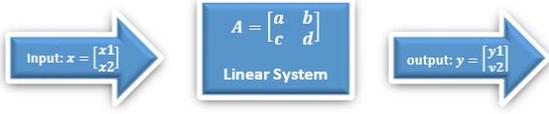

Fig. 1: Linear Transformation

Since an eigenvector of a matrix (e.g. $A$) only changes its magnitude when multiplied by the matrix, it indicates a direction in which $A$ would have maximum effect on the input [22]. In other words, any change in input which lies in the span of the eigenvector $v_{\lambda m}$ (corresponding to the highest eigenvalue, $\lambda_m$, of $A$), would bring maximum change in the output i.e. if $\Delta x = \alpha\, v_{\lambda m}$, where $\alpha$ is a scalar, then $\Delta y$ is maximized. The same concept can be applied to singular vectors of the matrix $A$ [23-24]. The singular value decomposition (SVD) of a matrix $A$ is defined as in (1)

$$A = U \Sigma V^T \quad (1)$$

Where columns of $U$ are called left singular vectors of $A$ (eigenvectors of $AA^T$) and rows of $V^T$ are called right singular vectors of $A$ (eigenvectors of $A^TA$). The eigenvalues of $AA^T$ are the singular values of $A$.

### B. Optimal Voltage Control

Using fast decoupled load flow assumptions, the changes in network bus voltage magnitudes can be approximated by (2)

$$\begin{bmatrix}\Delta V_{PV}\\ \Delta V_{PQ}\end{bmatrix} = [S_{VQ}]\begin{bmatrix}\Delta Q_{PV}\\ \Delta Q_{PQ}\end{bmatrix} \quad (2)$$

$PV$ is representing voltage controlled buses and $PQ$ is representing load buses. $S_{VQ}$ is $Q$-$V$ sensitivity matrix of the network and it can be decomposed as in (2a)

$$\begin{bmatrix}\Delta V_{PV}\\ \Delta V_{PQ}\end{bmatrix} = \begin{bmatrix}S_{11} & S_{12}\\ S_{21} & S_{22}\end{bmatrix}\begin{bmatrix}\Delta Q_{PV}\\ \Delta Q_{PQ}\end{bmatrix} \quad (2a)$$

where

$S_{11}$ = sensitivity of $\Delta V_{PV}$ on $\Delta Q_{PV}$
$S_{12}$ = sensitivity of $\Delta V_{PV}$ on $\Delta Q_{PQ}$
$S_{21}$ = sensitivity of $\Delta V_{PQ}$ on $\Delta Q_{PV}$
$S_{22}$ = sensitivity of $\Delta V_{PQ}$ on $\Delta Q_{PQ}$

Equation (2a) can be solved to get (3)

$$\Delta V_{PQ} = S_{21} S_{11}^{-1} \Delta V_{PV} + D \quad (3)$$

Here $D = (S_{22} - S_{21}S_{11}^{-1}S_{12})\Delta Q_{PQ}$, is considered as disturbance because we have no control over reactive power demand. Equation (3) gives us relation between load bus and source bus voltages. It can be used to control load bus voltages by adjusting voltages of $PV$ buses.

The objective of the controller is to find optimal input ($\Delta V_{PV}$) that would have maximum effect on the output ($\Delta V_{PQ}$). The objective function $J$ can be described by (4)

$$\max_{\Delta V_{PV}} J = \Delta V_{PQ}^T\, M\, \Delta V_{PQ} \quad (4)$$

Where $M$ is weight matrix to select the load bus we want to control or change more. Using (3), $J$ can be approximated (neglecting D) as in (5)

$$J = \Delta V_{PV}^T\, N\, \Delta V_{PV} \quad (5)$$

Where $N = S_{11}^{-T} S_{21}^T M\, S_{21} S_{11}^{-1}$. Objective function $J$ would be maximized if we select $\Delta V_{PV}$ in the span of that left singular vector of $N$ which corresponds to its highest singular value (i.e. $\sigma_1^2$). Using singular value decomposition of $N$ (i.e. $N = U \Sigma V^T$), (5) can be modified as

$$J = \Delta V_{PV}^T\, U \Sigma V^T\, \Delta V_{PV} \quad (6)$$

If $\Delta V_{PV} = \alpha u_1$, where $u_1$ is the left singular vector corresponding to the highest singular value of $N$, then performance index $J$ would be maximized and (6) would be modified as in (7)

$$J_{max} = \alpha^2 \sigma_1^2 \quad (7)$$

Scalar constant $\alpha$ is the design parameter in this control problem which would give required change in voltage of the load bus being controlled. In this approach, we are controlling one load bus voltage (having least deviation from reference voltage i.e. 1 pu) at a time using output weight matrix $M$. We call this *controlled bus (CB)* in our algorithm. Using (4) and (7), we can get (8)

$$J_{max} = \Delta V_{PQ}^T\, M\, \Delta V_{PQ} = \Delta V_{CB}^2 = \alpha^2 \sigma_1^2 \quad (8)$$

Design parameter $\alpha$ can be calculated using (8)

$$\alpha = \frac{\Delta V_{CB}}{\sigma_1} \quad (9)$$

### III. PROPOSED OPTIMAL VOLTAGE CONTROL ALGORITHM

In this paper, all buses whose voltages are outside normal voltage limits are called *critical buses*. The proposed optimal voltage control algorithm can be described by following steps:

1. Initialize $\Delta V_{PV} = \mathbf{0}$ (zero vector). Solve power flow and identify critical buses. If there are no critical buses, then return $\Delta V_{PV,\, required} = \Delta V_{PV}$ and terminate the algorithm.

2. Select the critical bus as *controlled bus* which has least deviation from the reference voltage (i.e. 1 pu).

3. Compute $\alpha$ which would make control bus voltage ($V_{CB}$) equal to reference voltage (1 pu).

4. Find $\Delta V_{PV}$ using (9) that would give required change in $V_{CB}$ (i.e. $\Delta V_{CB}$ in (8))

$$\Delta V_{PV} = \alpha u_1 \quad (9)$$

5. Update voltages of the voltage controlled buses ($V_{PV,\, new}$) using (10) and return to step 1.

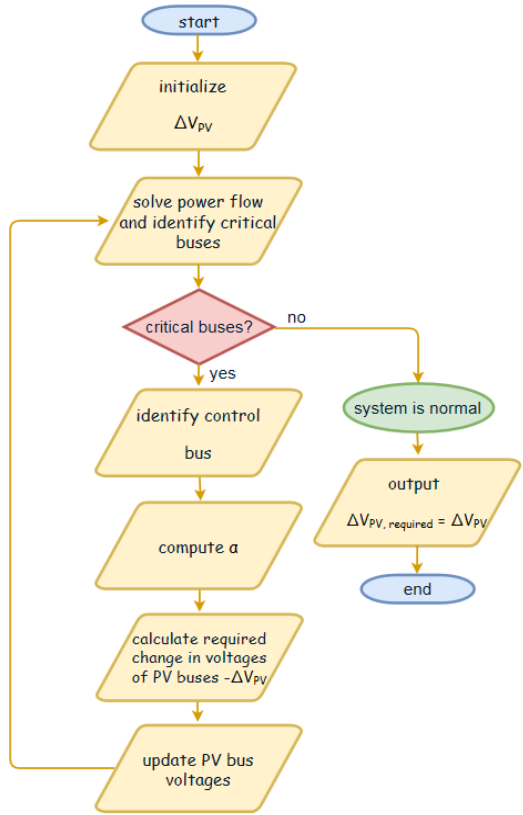

Fig. 2: Optimal Voltage Control Algorithm Flowchart

$$V_{PV,new} = V_{PV,old} + \Delta V_{PV} \quad (10)$$

Note: Instead of running power flow in each iteration, we can also use (11) to approximate $\Delta V_{PQ, new}$ and identify critical buses.

$$\Delta V_{PQ,new} \simeq S_{21} S_{11}^{-1} \Delta V_{PV,new} \quad (11)$$

The control flow diagram of the proposed voltage control algorithm is shown in Fig. 2.

## IV. CASE STUDIES

The proposed optimal voltage control algorithm was tested on three *IEEE* test systems, namely a 9 bus, a 14 bus and a 30 bus system respectively [25]. In every case, we randomly added some disturbances (inductive or capacitive) into the original system to make one or more bus voltages go out of limits and then applied this control algorithm to bring them back within normal limits. All simulations results were obtained using MATLAB and MATPOWER 6.0 [26].

### A. IEEE 9 Bus System

This system has three (bus 1-3) voltage controlled buses (including slack bus) and six (bus 4-9) load buses. The active power demand is 315 *MW* while reactive power demand is 115 *MVAR*. All voltages are within limits (0.9 to 1.1 pu) under normal operating conditions. The system diagram is shown in Fig. 3. We added a disturbance on bus 9 (70 *MVAR*) and its voltage came down to 0.8853 pu. After applying voltage control algorithm, its voltage was brought back to 0.9934 pu in one iteration (because there were no other voltage violations on load buses). Bus voltages before and after control are given in Table I. Change in voltages after applying voltage control algorithm are shown in Fig. 4. It is evident from the figure that sensitivity is inherent in this voltage control approach. Bus 9 is most sensitive to generator 1, and hence maximum change in bus 1 voltage, while least sensitive to generator 3 and hence minimum change in bus 3 voltage.

### B. IEEE 14 Bus System

This system has five (bus 1-3, 6, 8) voltage controlled buses (including slack bus) and nine (bus 4, 5, 7, 9-14) load buses. Active power demand is 259 *MW* and reactive power demand is 73.5 *MVAR*. All voltages are within limits under normal conditions. The system diagram is shown in Fig. 5. We applied a disturbance (-46.4 *MVAR*) on bus 10 and its voltage increased from 1.051 pu to 1.112 pu. After applying proposed voltage control algorithm, it came back within normal range in two iterations. In first iteration, bus 10 was controlled to bring it back to reference voltage (1 pu) but it pushed bus 12 below 0.9 pu (i.e. 0.899 pu). In second iteration, bus 12 was controlled and all buses remain within normal limits. Bus voltages before and after control for both iterations as well as change in voltages after applying voltage control algorithm has been shown in Fig. 6.

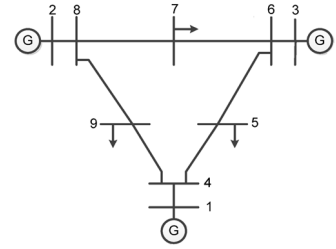

Fig. 3: *IEEE* – 9 bus System

TABLE I. Bus Voltages before and after Control (*IEEE* – 9bus)

| Bus | Voltage (pu) | |
|---|---|---|
| | Before | After |
| 1 | 1.000 | 1.100 |
| 2 | 1.000 | 1.086 |
| 3 | 1.000 | 1.040 |
| 4 | 0.960 | 1.060 |
| 5 | 0.954 | 1.045 |
| 6 | 0.995 | 1.057 |
| 7 | 0.971 | 1.051 |
| 8 | 0.978 | 1.066 |
| 9 | 0.885 | 0.993 |

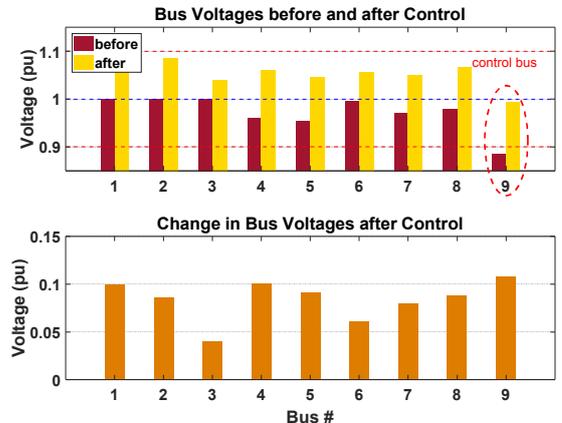

Fig. 4: Bus Voltages before and after Control (*IEEE* – 9bus)

We might face a situation in which we have two (or more) conflicting buses (i.e. one bus needs to decrease the *PV* bus voltage while the other wants to increase the same *PV* bus voltage) as critical buses. In this situation, the number of iterations would increase because we are controlling one bus at a time, but eventually system would come within normal voltage limits if there exist a solution. This problem can be solved by controlling multiple buses simultaneously. One such situation was studied by adding multiple disturbances into the system (bus 7 and 14). A conflict was observed between bus 7 and bus 14 because bus 7 was violating upper limit while bus 14 was violating lower limit.

The solution was obtained after five iterations. Bus voltages after each iteration are given in Table II while voltages for both buses in each iteration are shown in Fig. 7 along with the change in voltages after applying control algorithm (first iteration is representing voltages before control in the figure).

TABLE II. Bus Voltages before and after Control (*IEEE* – 14bus)

| Bus | Voltage (pu) | | | | | |
|---|---|---|---|---|---|---|
| | *Before* | *Iter. 1* | *Iter. 2* | *Iter. 3* | *Iter. 4* | *Iter. 5* |
| 1 | 1.000 | 1.013 | 0.989 | 1.004 | 0.987 | 1.000 |
| 2 | 1.000 | 1.041 | 0.964 | 1.013 | 0.959 | 1.001 |
| 3 | 1.000 | 1.025 | 0.978 | 1.008 | 0.975 | 1.001 |
| 4 | 0.984 | 1.033 | 0.953 | 1.009 | 0.955 | 1.001 |
| 5 | 0.976 | 1.026 | 0.950 | 1.005 | 0.953 | 0.998 |
| 6 | 1.000 | 1.100 | 0.998 | 1.1 | 1.028 | 1.100 |
| 7 | 1.074 | 1.145 | 1.022 | 1.102 | 1.028 | 1.094 |
| 8 | 1.000 | 1.084 | 0.900 | 1 | 0.900 | 0.988 |
| 9 | 1.004 | 1.086 | 0.963 | 1.054 | 0.976 | 1.047 |
| 10 | 0.996 | 1.082 | 0.961 | 1.054 | 0.977 | 1.049 |
| 11 | 0.994 | 1.087 | 0.976 | 1.073 | 0.999 | 1.071 |
| 12 | 0.967 | 1.070 | 0.962 | 1.067 | 0.992 | 1.066 |
| 13 | 0.948 | 1.051 | 0.939 | 1.046 | 0.968 | 1.044 |
| 14 | 0.828 | 0.937 | 0.794 | 0.912 | 0.819 | 0.908 |

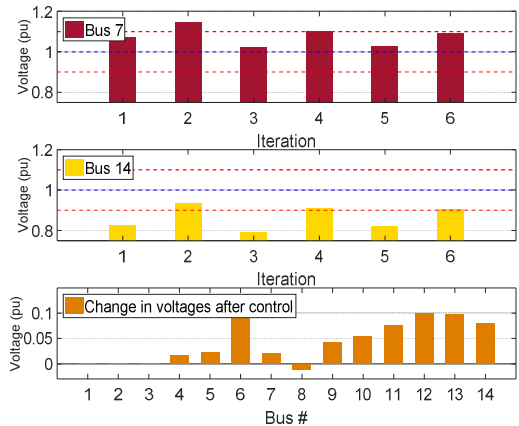

Fig. 7: Bus 7 and Bus 14 Voltages in each iteration (*IEEE* – 14bus)

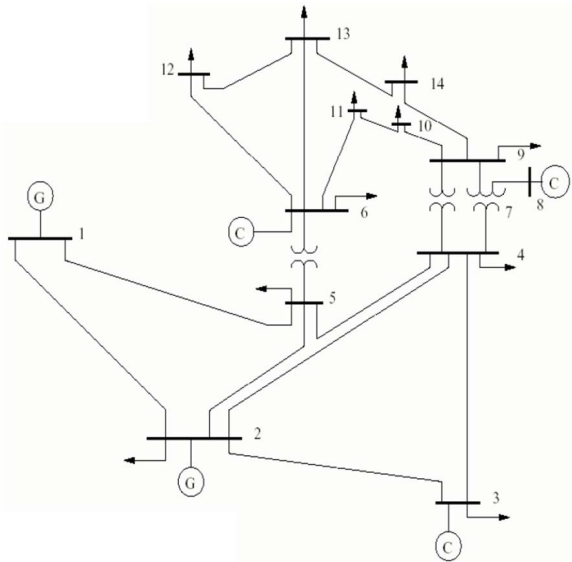

Fig. 5: *IEEE* – 14 bus System

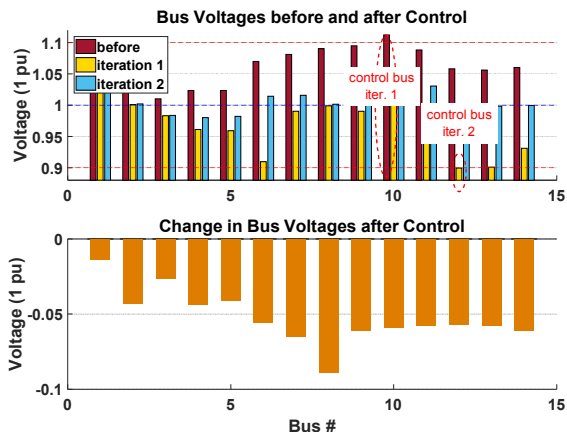

Fig. 6: Bus Voltages before and after Control (*IEEE* – 14bus)

### C. IEEE 30 Bus System

This system has 6 voltage controlled buses (bus 1, 2, 13, 22, 23, 27) and 24 load buses. Active power demand is 189.2 *MW* and reactive power demand is 107.2 *MVAR*. All voltages are within normal limits under normal operating conditions. The system diagram is shown in Fig. 8. We added disturbances on bus 8 (90 *MVAR*) and bus 25 (-100 *MVAR*) which resulted in voltage violations on bus 8 (.878 pu) and bus 25 (1.116 pu). After applying voltage control algorithm, all voltages came back within normal limits in two iterations. Bus voltages for each iteration are given in Table III. While voltages in each iteration as well as change in voltages after applying voltage control algorithm are shown in Fig. 9.

### V. PERFORMANCE COMPARISON

We compared our optimal voltage control approach (*OVC*) with sensitivity based voltage control approach (*SVC*) which uses sensitivities of *PQ* bus voltages on *PV* bus voltages (changing *PV* bus voltage to which control bus is most sensitive). The performance index described in (4) that it is being maximized was computed for comparison. The performance

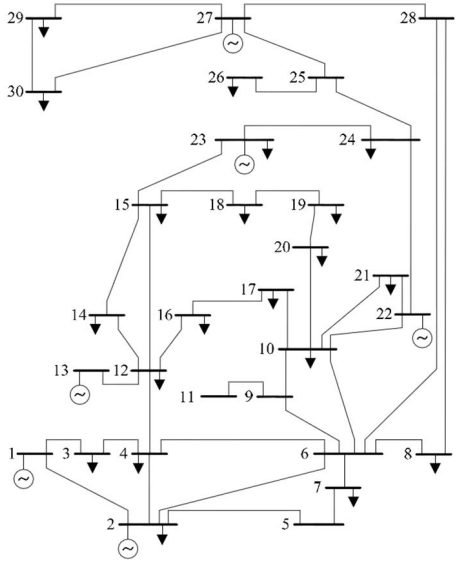

Fig. 8: *IEEE* – 30 bus System

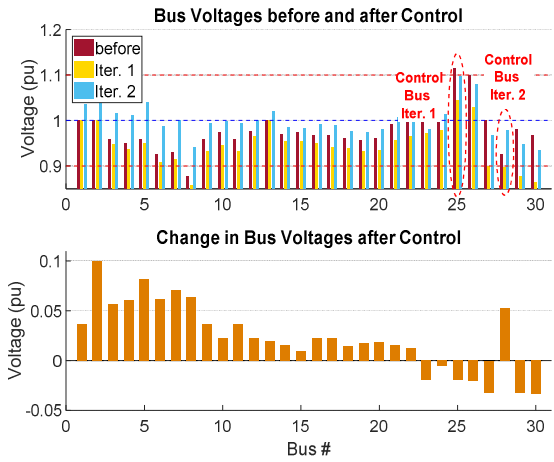

Fig. 9: Bus voltages in each iteration (IEEE – 30bus)

TABLE III. Bus voltages in each iteration (*IEEE* – 30bus)

| Bus | Voltage (pu) | | | Bus | Before | Iter. 1 | Iter. 2 |
|---|---|---|---|---|---|---|---|
| | *Before* | *Iter. 1* | *Iter. 2* | | | | |
| 1 | 1.000 | 1.000 | 1.037 | 16 | 0.968 | 0.948 | 0.991 |
| 2 | 1.000 | 1.000 | 1.100 | 17 | 0.967 | 0.941 | 0.990 |
| 3 | 0.957 | 0.947 | 1.014 | 18 | 0.961 | 0.938 | 0.976 |
| 4 | 0.949 | 0.936 | 1.010 | 19 | 0.957 | 0.932 | 0.974 |
| 5 | 0.959 | 0.950 | 1.041 | 20 | 0.961 | 0.934 | 0.979 |
| 6 | 0.926 | 0.908 | 0.988 | 21 | 0.991 | 0.957 | 1.006 |
| 7 | 0.929 | 0.915 | 1.000 | 22 | 1.000 | 0.965 | 1.013 |
| 8 | 0.878 | 0.857 | 0.941 | 23 | 1.000 | 0.972 | 0.980 |
| 9 | 0.958 | 0.932 | 0.994 | 24 | 1.020 | 0.977 | 1.014 |
| 10 | 0.975 | 0.945 | 0.998 | 25 | 1.116 | 1.045 | 1.096 |
| 11 | 0.958 | 0.932 | 0.994 | 26 | 1.100 | 1.028 | 1.079 |
| 12 | 0.977 | 0.964 | 1.000 | 27 | 1.000 | 0.900 | 0.968 |
| 13 | 1.000 | 1.000 | 1.019 | 28 | 0.925 | 0.898 | 0.977 |
| 14 | 0.969 | 0.953 | 0.985 | 29 | 0.980 | 0.877 | 0.947 |
| 15 | 0.974 | 0.955 | 0.984 | 30 | 0.968 | 0.864 | 0.935 |

TABLE IV. Comparison of Performance Indices for *OVC* and *SVC*

| *Case Study* | OVC | SVC |
|---|---|---|
| IEEE 9 bus | 0.0117 | 0.0042 |
| IEEE 14 bus | 0.0119 | 0.0068 |
| IEEE 30 bus | 0.0052 | 0.0038 |

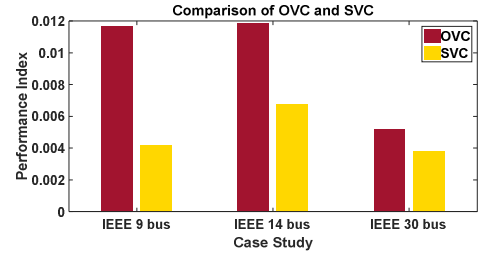

Fig. 10: Comparison of OVC and SVC

index (for one iteration) for both cases are given in Table IV and shown in Fig. 10.

It is evident from the results that the proposed optimal voltage control (*OVC*) approach is giving better performance index in all cases and hence better approach compared to sensitivity based voltage control (*SVC*).

## VI. CONCLUSION AND FUTURE WORK

In this paper, an optimal voltage control was proposed using the Fast Decoupled Load Flow (FDLF) jacobian matrix. The proposed algorithm was tested on different standard *IEEE* systems and satisfactory results were obtained. In case of no conflict, the maximum number of iterations in which the algorithm converges is either equal to or less than the total number of critical buses at first iteration. The drawback of this approach is that it controls only one bus at a time resulting in more number of iterations in cases where we have conflicting buses as critical buses. This problem can be addressed by simultaneously considering multiple buses for control. These issues are being investigated at this time.